\documentclass[english,preprintnumbers,amsmath,amssymb,twocolumn,reprint,nofootinbib]{revtex4-1}
\usepackage[T1]{fontenc}
\usepackage[latin9]{inputenc}
\usepackage{color}
\usepackage{float}
\usepackage{amsbsy}
\usepackage{amstext}
\usepackage{graphicx}
\usepackage{esint}

\usepackage{amsmath}

\makeatletter

\@ifundefined{textcolor}{}
{%
 \definecolor{BLACK}{gray}{0}
 \definecolor{WHITE}{gray}{1}
 \definecolor{RED}{rgb}{1,0,0}
 \definecolor{GREEN}{rgb}{0,1,0}
 \definecolor{BLUE}{rgb}{0,0,1}
 \definecolor{CYAN}{cmyk}{1,0,0,0}
 \definecolor{MAGENTA}{cmyk}{0,1,0,0}
 \definecolor{YELLOW}{cmyk}{0,0,1,0}
 }

\usepackage{dcolumn}% Align table columns on decimal point
\usepackage{bm}% bold math
\usepackage{array}

\def\ket{\rangle}
\def\bra{\langle}

\def\epsd{\epsilon_\textrm{d}}

\def\im{\mathop{\textrm{Im}}}

\newcommand{\beq}{\begin{equation}}
\newcommand{\eeq}{\end{equation}}
\newcommand{\beqa}{\begin{eqnarray}}
\newcommand{\eeqa}{\end{eqnarray}}

\makeatother

\usepackage{babel}

\begin{document}

\title{Non-Markovian dynamics revealed at a bound state in continuum}
%\title{Localization transitions and non-exponential decay: physicist's proof and an experimental proposal}

\author{Savannah Garmon}
\email{sgarmon@p.s.osakafu-u.ac.jp}
\author{Kenichi Noba}
\affiliation{Department of Physical Science,
Osaka Prefecture University,
Gakuen-cho 1-1, Sakai 599-8531, Japan}

\author{Gonzalo Ordonez}
\affiliation{Department of Physics and Astronomy, Butler University,
Gallahue Hall, 4600 Sunset Ave.,
Indianapolis, Indiana 46208, USA}

\author{Dvira Segal}
\affiliation{Chemical Physics Theory Group, Department of Chemistry,
University of Toronto, 80 Saint George St., Toronto, 
Ontario, M5S 3H6, Canada}

\begin{abstract}
We propose a methodical approach to controlling and enhancing deviations from exponential decay in quantum and optical systems by exploiting recent progress surrounding another subtle effect: the bound states in continuum, which have been observed in optical waveguide array experiments within this past decade.  Specifically, we show that by populating an initial state orthogonal to that of the bound state in continuum, it is possible to engineer system parameters for which the usual exponential decay process is suppressed in favor of inverse power law dynamics and coherent effects that typically would be extremely difficult to detect in experiment.  We demonstrate our method using a model based on an optical waveguide array experiment, and further show that the method is robust even in the face of significant detuning from the precise location of the bound state in continuum.
\end{abstract}

%\received{October 15, 2016}
%\revised{April 18, 2017}

\date{\today}
\maketitle

%\section{Introduction}
%\label{sec:intro}

A bound state in continuum (BIC) represents a localized eigenmode with energy eigenvalue that, counter-intuitively, resides directly within the scattering continuum of a given physical system.  Although the existence of such modes were first predicted in 1929 \cite{BIC_1929}, the phenomenon is so delicate that they were not observed until much more recently \cite{BIC_review}; for example, in optical waveguide array experiments \cite{BIC_opt_expt0,BIC_opt_expt1,BIC_opt_exptA,BIC_opt_expt2}.  Lasing action has also recently been reported for a cavity supporting BICs \cite{BIC_lasing}.  In this work, we propose to apply these recent technical advances in optical control of the BIC to the study of another often elusive phenomenon: long-time non-exponential decay.

In many familiar circumstances, such as atomic relaxation, we tend to think of quantum decay as essentially an exponential process.  More precisely, exponential decay tends to manifest when an unstable eigenmode (such as an excited atomic level) is resonant with an energy continuum (environmental reservoir, such as the electromagnetic vacuum) to which it is coupled.  
However, it can be shown that in fact all quantum systems follow non-exponential dynamics on very short and extremely long timescales.  These deviations occur as a direct result of the existence of at least one threshold on the energy continuum in such systems \cite{Khalfin,Winter61,Fonda,MDCG06,Muga_review2,GPSS13,OH17,CS18}.  While these effects are ubiquitous in quantum systems, they are unfortunately quite difficult to detect under ordinary circumstances and hence have been measured only in a small handful of 
experiments \cite{short_time_expt,zeno_expt,zeno_expt_latt,long_time_expt,KurizkiZeno,ZenoSC}.
The short-time deviation, which can give rise to both decelerated \cite{SudarshanZeno} and accelerated \cite{antiZeno} decay under frequent system observations \cite{SudarshanZeno,zeno_expt,KurizkiZeno,ZenoSC} or modulation of the environmental coupling \cite{zeno_expt_latt}, requires ultra-precision to detect that is often difficult to achieve in the lab.
Ref. \cite{XCLYG14} uses the properties of a BIC to study these short-time effects.

The long-time deviation, meanwhile, has proven even more challenging \cite{long_time_expt}.  The difficulty originates in that the effect usually does not appear until many lifetimes of the exponential decay have passed, by which time the survival probability is so depleted that it is rendered undetectable.
%%%%%
A handful of theory papers have suggested special circumstances to enhance the long-time effect; these mostly require an initially prepared state near the threshold, usually combined with other conditions \cite{KKS94,JQ94,LNNB00,Jittoh05,GCV06,Longhi06,DBP08,GPSS13,GO17}. See also the recent experiment \cite{Krinner18}.

In this paper we take advantage of the simple geometric shape of the BIC to present a qualitatively different and more easily generalized scheme by which the long-time deviation can be enhanced.  
While it is clear from the outset that the usual exponential decay associated with the resonance is suppressed when the BIC condition is satisfied, if one were to directly populate the BIC itself then one would observe a simple stable evolution, as the BIC is of course an eigenstate of the Hamiltonian.
However, we show that by populating a state that is orthogonal to the BIC we can take advantage of the suppression of the exponential effect while avoiding the stability associated with the BIC itself.  The non-exponential dynamics can then drive the evolution on all timescales.  What's more, we demonstrate in our example below that the exponential effect can be dramatically suppressed even with significant detuning from the BIC, although the choice of BIC-orthogonal initial state is still essential.  %\footnote{\textcolor{blue}{We emphasize that, unlike Refs \cite{zeno_expt,SudarshanZeno,antiZeno,zeno_expt_latt,KurizkiZeno,ZenoSC,XCLYG14}, our method does not rely on frequent system interruptions to observe non-exponential effects.} }

We illustrate our method relying on a simple tight-binding model that can be viewed analogously
to one of the previously mentioned optical waveguide array experiments. %\footnote{\textcolor{blue}{More precisely, the time evolution in the the quantum picture for model Eq. \ref{ham.n2.n} would compare to propagation along the optical waveguides.}  }
%Specifically, we consider a semi-infinite tight-binding chain with a side-coupled impurity
Our Hamiltonian is written
\beqa
  H = \epsd | d \ket \bra d | %+ \sum_{n=1}^\infty \epsilon_\mu  | n \ket \bra n |
  			& - & J \sum_{n=1}^\infty \left( | n \ket \bra n+1 | + | n+1 \ket \bra n | \right)
 		  	\nonumber	\\
  	& &	
		- g \left( | d \ket \bra 2 | + | 2 \ket \bra d | \right)
	,
\label{ham.n2.n}
\eeqa
in which the second term represents the semi-infinite array with nearest-neighbor hopping parameter $-J$ and the chain is side-coupled at site $| 2 \ket$ to an ``impurity'' element $| d \ket$.
%Without loss of generality we can set the common chain-impurity chemical potential $\epsilon_\mu = 0$ from this point forward.
After we set the energy units according to $J=1$, the adjustable parameters in the system are the chain-impurity coupling $-g$ and the impurity energy level $\epsd$.
This model captures the essential features of the waveguide array experiment in Ref. \cite{BIC_opt_expt2} (see Ref. \cite{LonghiEPJ07} as well as \cite{Fukuta}), when we view time evolution in the present context analogously to longitudinal propagation within the waveguides.
This model can be partially diagonalized by introducing a half-range Fourier series on the chain according to $| n \ket = \sqrt{\frac{2}{\pi}} \int_0^\pi dk \; \sin nk | k \ket$, after which we have
\beqa
  H = \epsd | d \ket \bra d | & + & \int_0^\pi dk \; E_k | k \ket \bra k | 
  			\nonumber	\\
		& & + g \int_0^\pi dk \; V_k \left( | d \ket \bra k | + | k \ket \bra d | \right)
\label{ham.n2.k}
\eeqa
where $V_k = - \sqrt{\frac{2}{\pi}} \sin 2 k$ and the continuum is given by $E_k = - 2 J \cos k$ over 
$k \in [0,\pi]$.  Note from here we will measure the energy in units of $J=1$.

The discrete spectrum for this model can be obtained, for example, from the resolvent operator
\beq
  %G_{dd}(z) = 
  \bra d | \frac{1}{z - H} | d \ket \; = \;
  	\frac{1}{z - \epsd - \Sigma (z)}
\label{n2.res.dd}
\eeq
in which the self-energy function $\Sigma (z) = g^2 \int_0^\pi dk \frac{| V_k |^2}{z - E_k}$
%The self-energy function $\Sigma(z)$ here is given by
%%
%\beq
%  \Sigma (z)
%  	= g^2 \int_0^\pi \frac{| V_k |^2}{z - E_k} dk
	%%
%	= \frac{z g^2}{2 J^4} \left[ z^2 - 2J^2 - z \sqrt{z^2 - 4 J^2} \right]
%	= \frac{z g^2}{2} \left[ z^2 - 2 - z \sqrt{z^2 - 4} \right]
%\label{n2.sigma.z.defn}
%\eeq
%%
is evaluated as
\beq
  \Sigma (z)
%  	= g^2 \int_0^\pi \frac{| V_k |^2}{z - E_k} dk
	%%
%	= \frac{z g^2}{2 J^4} \left[ z^2 - 2J^2 - z \sqrt{z^2 - 4 J^2} \right]
	= \frac{z g^2}{2} \left[ z^2 - 2 - z \sqrt{z^2 - 4} \right]
\label{n2.sigma.z}
\eeq
%%
%in which the upper (lower) sign choice is for $z>0$ ($z<0$) in the first Riemann sheet of the complex $z$-plane (the sign designations are reversed upon analytic continuation into the second sheet).}
in the first Riemann sheet [see Ref. \cite{Supp} for discussion of the analytic properties of $\Sigma(z)$].
Notice that a pole occurs in Eq. (\ref{n2.res.dd}) at $z=0$ after choosing $\epsd = 0$; this is the BIC solution for this model, which resides directly at the center of the continuum $z \in [-2, 2]$ (defined by 
the range of $E_k$) and which takes the form
\beq
  | \psi_\textrm{BIC} \ket
%  	= \frac{1}{\sqrt{J^2 + g^2}} \left( J | d \ket - g | 1 \ket \right)
  	= \frac{1}{\sqrt{1 + g^2}} \left( | d \ket - g | 1 \ket \right)
	.
\label{n2.bic}
\eeq
We here emphasize that the BIC state can be understood as a resonance with vanishing decay width \cite{Fukuta,BIC_review,SH75,FW85,Robnik,ONK06,SBR06,Rotter_review,QBIC,BS08,Moiseyev_BIC,Reichl09,GurvitzBIC,ZBK12,MA14,Boretz14,GTC17}.  In this picture, the ordinary resonance represents a generalized eigenstate with complex energy eigenvalue, for which the imaginary part of the eigenvalue gives the exponential decay half-width.  When the BIC condition $\epsd = 0$ is fulfilled the imaginary part of this eigenvalue vanishes, yielding a bound state residing directly in the scattering continuum.  When $\epsd \neq 0$ the complex eigenvalue is restored and the exponential decay would generally be expected to reappear.  
%%%%Following sentence could be deleted for space:
%Meanwhile note that the BIC solution exists in the present model for any value of the coupling $g$ as long as $\epsd = 0$ is satisfied.
%%%%

\begin{figure}
%\hspace*{0.01\textwidth}
 \includegraphics[width=0.45\textwidth]{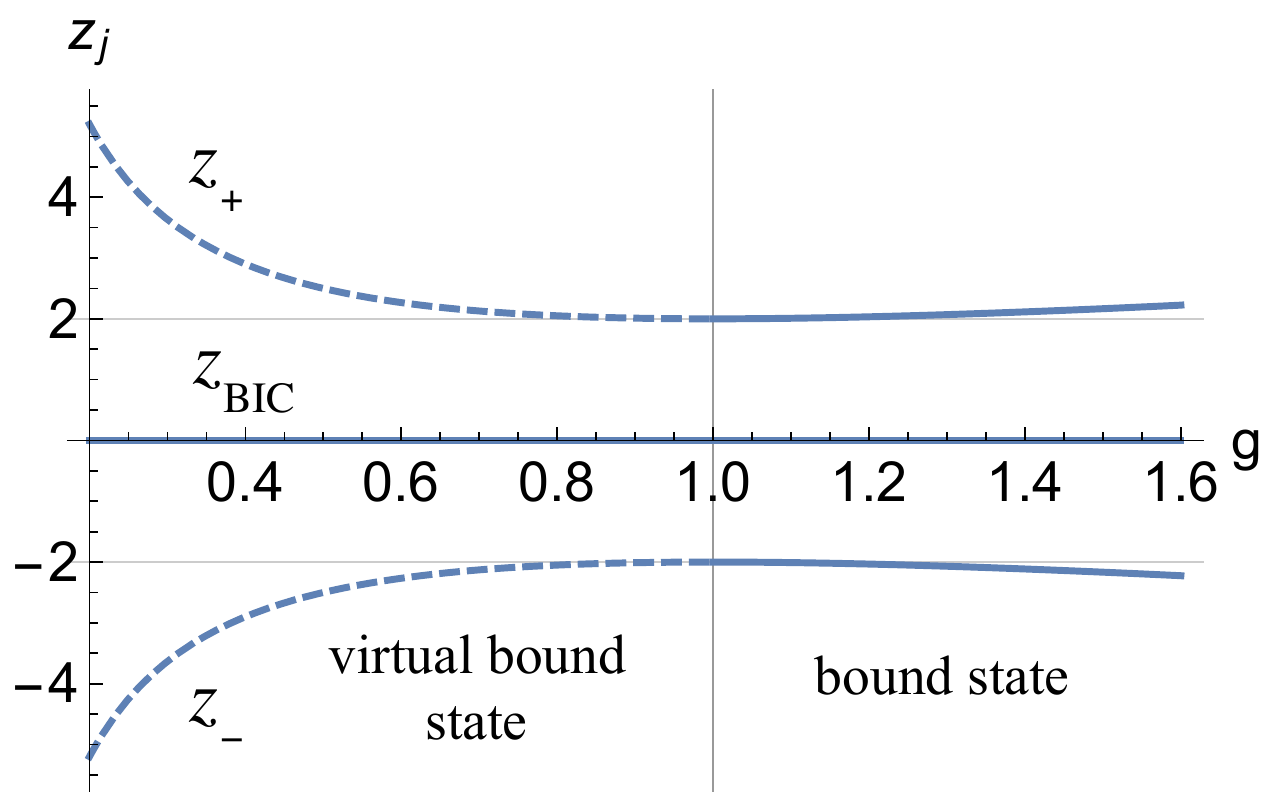}
\\
\vspace*{\baselineskip}
\caption{(color online) Discrete spectrum of our model as a function of $g$ in the case $\epsd = 0$.  The BIC appears at $z_\textrm{BIC} = 0$.  The other two solutions are virtual bound states (dashed curves) for $g < 1$; they become bound states for $g>1$. (Energy is measured in units of $J=1$ throughout the paper.)
}
\label{fig:0.spec} 
\end{figure}

It is easy to show that there exist two further solutions for the $\epsd = 0$ case with eigenvalues given by $z_\pm = \pm z_g$, in which
\beq
  z_g = g + \frac{1}{g}
  	.
\label{n2.zg}
\eeq
For $g>1$ these two solutions constitute localized bound states residing on the first Riemann sheet of the complex energy plane, while for $g < 1$ they transition to so-called virtual bound states (or anti-bound states), which are delocalized pseudo-states with real eigenvalue resting in the second sheet \cite{DBP08,GPSS13,Nussenzveig59,Hogreve,Moiseyev_NHQM,HO14},  see Fig \ref{fig:0.spec}.  
%Note that precisely such a delocalization transition is observed in the optical waveguide array experiment in Fig. 2(b) of Ref. \cite{BIC_opt_expt2}.
While the virtual bound states do not appear in the diagonalized Hamiltonian, they nevertheless have a similar influence on the long-time power law decay as do the bound states \cite{GPSS13}.    Specifically, we will show that the timescale characterizing the non-exponential decay is proportional to 
$\Delta_g^{-1}$, where
\beq
  \Delta_g
  	\equiv z_g - 2
	.
\label{n2.Deltag}
\eeq
is defined as the gap between either of the (virtual) bound state energies and the nearest band edge.
Note we will particularly focus on the $g \le 1$ portion of the parameter space as the absence of bound states here means that nothing inhibits the non-exponential decay.  (For comparison, we will also briefly discuss the $g > 1$ evolution.)

%%%
%%%

As previously discussed, if we were to consider the evolution of the BIC state itself, the initial state would simply remain occupied for all time as $| \psi_\textrm{BIC} \ket$ is an eigenstate of $H$ with energy eigenvalue $z=0$.  However, by instead choosing the (simplest) BIC-orthogonal state
\beq
  | \psi_\perp \ket
%  	= \frac{1}{\sqrt{J^2 + g^2}} \left( g | d \ket + J | 1 \ket \right)
  	= \frac{1}{\sqrt{1 + g^2}} \left( g | d \ket + | 1 \ket \right)
\label{n2.perp}
\eeq
as our initial state, we obtain complete non-exponential decay for any value $g \le 1$, as shown below \footnote{One might pause at the inclusion of the site $|1\ket$ that is technically part of the reservoir in this initial state.  However, $|1\ket$ could equivalently be viewed as a second impurity element \cite{Supp}. }.  
To analyze the evolution of $| \psi_\perp \ket$, we evaluate the survival probability $P_\perp (t) = |A_{\perp} (t)|^2$, in which the survival amplitude is given by
\beq
  A_{\perp} (t)
  	= \bra \psi_\perp | e^{-iHt} | \psi_\perp \ket
	= \frac{1}{2 \pi i} \int_{\mathcal{C}_E} e^{-izt} \bra \psi_\perp | \frac{1}{z - H} | \psi_\perp \ket \; dz
	.
\label{A.perp.defn}
\eeq
Here $\mathcal{C}_E$ is a counter-clockwise integration contour surrounding the real axis in the first Riemann sheet of the complex energy plane, which includes the branch cut along $z \in [-2,2]$ as well as any bound states.  
We can apply various methods to evaluate this integral, for example by directly computing the relevant matrix elements of the resolvent operator and integrating over these or by applying an expansion in terms of the eigenstates of the generalized discrete spectrum of the model as in Ref. \cite{HO14}.  By either method we obtain the following results.

\begin{figure*}
%\hspace*{0.01\textwidth}
 \includegraphics[width=0.85\textwidth]{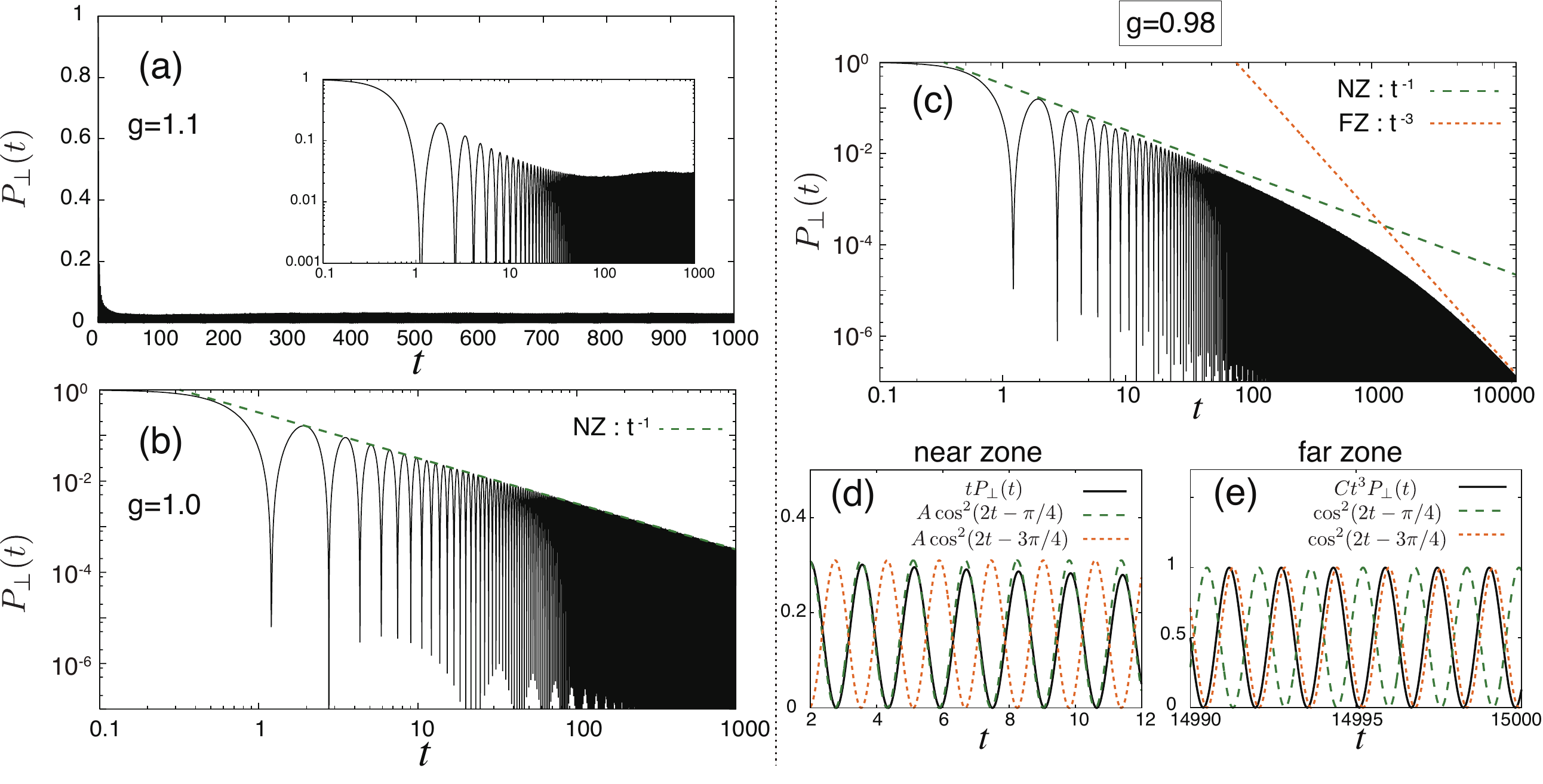}
\\
\vspace*{\baselineskip}
\caption{(color online) Numerical simulations for the survival probability of $|\psi_\perp \ket$ at time $t$ for $\epsd = 0$ and (a) $g=1.1$ (linear plot, inset: log-log plot), (b) $g=1.0$ (log-log plot), and (c, d, e) g=0.98 (c: log-log plot, d: early near zone close-up, and e: far zone close-up).  The green dashed (orange dotted) lines indicate the $1/t$ ($1/t^3$) %near (far) zone  
dynamics.
The numerical method is described in \cite{Supp}.
(Time $t$ is measured in units $1/J$ in which $J=1$.)
}
\label{fig:n2.surv.prob} 
\end{figure*}

For the case $g > 1$ there are two bound states included in the contour for Eq. (\ref{A.perp.defn}).  The survival amplitude in this case evaluates as
\beq
  A_\perp (t)
  	= \frac{g^2 - 1}{g^2} \cos z_g t + A_\textrm{br} (t)
	,
\label{A.perp.poles}
\eeq
in which the first term represents the combined contributions from the two bound states 
while 
%the contribution due to the integration along the branch cut is given by
%%
\beq
  A_\textrm{br} (t)
  	= \frac{1+g^2}{4 \pi i g^2} \int_{\mathcal{C}_\textrm{br}} dz e^{-izt} \frac{\sqrt{z^2 - 4}}{z^2 - z_g^2}
	.
\label{A.br.defn}
\eeq
is an integration along the contour $\mathcal{C}_\textrm{br}$ surrounding the branch cut in a counter-clockwise manner in the complex energy plane.
The decay in this case is non-exponential but incomplete due to the presence of the bound states \cite{KKS94,JQ94,LNNB00}.
%The presence of the bound states in this case results in incomplete decay, which tends to de-emphasize the non-exponential evolution associated with the branch cut.  
This can be seen for the case $g=1.1$ in Fig. \ref{fig:n2.surv.prob}(a).

%\begin{widetext}

%%%
%%%

%\end{widetext}

Meanwhile for the case $g \le 1$, the bound states have become virtual bound states and the evolution is now determined entirely by the non-Markovian branch cut contribution $A_\perp (t) = A_\textrm{br} (t)$.
%, which yields pure non-exponential decay for the entire evolution.  
We find that this expression yields two distinct time regions, in which the integral is most easily estimated by somewhat different methods. 
%These are the short/intermediate time region we call the {\it non-exponential near zone} and the asymptotic (long time) region we call the {\it non-exponential far zone}, similar to the convention introduced in Ref. \cite{GPSS13}.
First there is a short/intermediate time region, in which we first apply a fraction decomposition to the denominator of Eq. (\ref{A.br.defn}); this yields two simpler integrals, one associated with the upper virtual bound state and the other associated with the lower.
%These two integrals can then be evaluated using methods similar to those appearing in Apps. C and D of Ref. \cite{GO17}, which yields
As outlined in \cite{Supp}, these two integrals can be evaluated in terms of Bessel functions by methods similar to those used in Ref. \cite{GO17}, which yields
\beq
  A_\textrm{br} (t)
  	\approx \frac{1}{g} J_0 (2 t) - \frac{1-g}{g} \cos 2 t
	;
\label{n2.A.early}
\eeq
this expression holds for all $t \ll T_\Delta$ where $T_\Delta$ is written as 
$T_\Delta = 1/\Delta_g = g/(1-g)^2$ in terms of the energy gap between the virtual bound states and their respective nearby band edges.  On the earliest timescale $t \ll T_Z$ with $T_Z = 1$, this expression yields the usual short time parabolic dynamics $P_Z (t) \approx 1 - C t^2$, in which 
$C = (g + g^2 + g^3 - 1)/g^2$.

Then, in the intermediate time region $T_Z \ll t \ll T_\Delta$, we can approximate the Bessel function in the first term of Eq. (\ref{n2.A.early}) to write
\beq
  A_\textrm{NZ} (t)
  	\approx \frac{\cos (2t - \pi / 4)}{g \sqrt{\pi t}} - \frac{1-g}{g} \cos 2 t
	.
\label{n2.A.NZ}
\eeq
We refer to this time region including characteristic $1/t$ decay as the {\it non-exponential near zone} (NZ) \cite{GPSS13}, which we can roughly think of as having replaced the usual exponential decay regime.
For values $g \lesssim 1$ fairly close to the $g=1$ localization transition, the first term in Eq. (\ref{n2.A.NZ}) tends to dominate the evolution early in the near zone, while the second term provides only a small correction.
Estimating the evolution in this case %from the first term of Eq. (\ref{n2.A.NZ}) alone 
yields 
\beq
  P_\textrm{NZ,early} (t)
  	\approx \frac{ \cos^2 \left( 2 t - \pi/4 \right)}{\pi g^2 t}
	,
\label{n2.P.NZ}
\eeq
which can be seen for the case $g=0.98$ in Fig. \ref{fig:n2.surv.prob}(c,d).
As we move later into the near zone, the first term decays sufficiently so that the second term becomes non-negligible; we can estimate this as when the second term is about 10\% of the first, which gives $t = T_\textrm{VR} \sim 1 / \left[100 \pi (1-g)^2 \right] = 1/ \left[100 \pi g \right] * T_\Delta$.  This implies we should be fairly close to the transition point $g=1$ to observe the pure $1/t$ dynamics.  For example, in the case $g=0.98$ shown in Fig. \ref{fig:n2.surv.prob}(d), we can already see a small influence from the second term of Eq. (\ref{n2.A.NZ}) around 
$t \gtrsim T_\textrm{VR} \approx 8.0$ as the last three visible oscillation cycles show a slight deviation from the Eq. (\ref{n2.P.NZ}) prediction, which the second term of Eq. (\ref{n2.A.NZ}) [not shown] captures very well.
We will return to the physical interpretation of this second term momentarily.

Next appears the asymptotic time region $T_\Delta \ll t$ during which the dynamics are instead described by a $1/t^3$ power law decay.  To show this, we return to the (exact) integral expression for the survival amplitude appearing in Eq. (\ref{A.br.defn}) and instead proceed by deforming the contour $\mathcal{C}_\textrm{br}$ surrounding the branch cut by dragging it out to infinity in the lower half of the complex energy plane, as described in \cite{Supp}.
%After this, the only remaining non-vanishing portions of the integration are the contours $A_\mp (t)$ running from the two branch points out to infinity in the lower half plane. Applying an integration variable transform $s \equiv it (z \pm 2)$ allows us to write these two contributions as
%%
%\beq
 % A_\mp (t)
%  	  = \frac{i \left(1+g^2\right) e^{\pm 2it }}{2 \pi g^2 t^2} 
%	  		\int_{0}^{\infty} ds e^{-s} \frac{\sqrt{s^2 \mp 4ist}}
%					{\mp \Delta_g \left( 2 + z_g \right) + 4i\frac{s}{t} \mp \frac{s^2}{t^2}}
%	.
%\label{A.br.mp.2}
%\eeq
%%
%For $t$ large in this asymptotic region, we can neglect the second and third terms in the denominator of the integral, so that finally we obtain
Following this procedure, we obtain
\beq
  P_\textrm{FZ} (t) %\approx | A_- + A_+ |^2 
  	\approx \frac{\left(1 + g^2 \right)^2 \cos^2 \left( 2 t - 3 \pi/4 \right)}
					{\pi g^4 \Delta_g^2 \left( 2 + z_g \right)^2 t^3}
	,
\label{n2.P.FZ}
\eeq
with the characteristic $1/t^3$ decay that is typical of odd dimensional systems on long timescales \cite{Sudarshan,Muga95,GPSS13,GCMM95,Cavalcanti98,Zueco17}.  
We refer to this as the {\it non-exponential far zone} (FZ).
The far zone dynamics can be seen for the $g=0.98$ case in Fig. \ref{fig:n2.surv.prob}(c,e).

We emphasize three further points about these results as follows.
%%Oscillations and phase shift
First, we draw attention more carefully to the occurrence of oscillations in both time zones, which are due to interference between the contributions from the two band edges.  These contributions are equally weighted because the BIC occurs at the center of the continuum band in the present case.  Notice further that a $\pi/2$ phase shift occurs between the early near zone result Eq. (\ref{n2.P.NZ}) and the far zone Eq. (\ref{n2.P.FZ}).
These oscillations and the resulting phase shift are highlighted in Fig. \ref{fig:n2.surv.prob}(d,e).  While similar oscillations have been previously predicted in the far zone \cite{Longhi06,GO17,Zueco17}, we believe the near zone oscillations as well as the resulting phase shift are new --- indeed, outside of our choice for the initial state, these would almost certainly be obscured by the exponential decay.
%%virtual Rabi effect
Second, we return our attention to the second term of Eq. (\ref{n2.A.NZ}), which becomes relatively more pronounced later in the near zone; however, counterintuitively perhaps, it vanishes in the far zone\footnote{The reason for this is discussed in pp. 21-22 of Ref. \cite{HO14}.}.
Notice this term takes the form of a Rabi-like oscillation between the band edges at $z=\pm 2$.  We refer to this effect as a {\it virtual Rabi oscillation}, which is intended to reflect its transient nature.  A further interesting point is that the virtual Rabi oscillation plays a role in facilitating the phase shift from the early near zone into the far zone \cite{Supp}.  
%For example, it easy to show that around $t = g/4\pi * T_\Delta$ in the near zone evolution the coefficient of the second term in Eq. (\ref{n2.A.NZ}) has about half the magnitude of that of the first term; in this case, the effective phase of the two combined terms can be approximated as 
%$\phi_{1/2} = \arctan (\sqrt{2}/(\sqrt{2} -1)) \approx 0.4093 \pi$, which indeed satisfies $\pi/4 < \phi_{1/2} < 3\pi/4$.

Third, notice that when we are directly at the localization transition at $g=1$, the second term in Eq. (\ref{n2.A.NZ}) vanishes.  Further, since the key timescale $T_\Delta$ is inversely proportional to $\Delta_g$, as we approach $g=1$ from below the energy gap $\Delta_g$ closes and $T_\Delta$ diverges.  Hence, in this case, Eq. (\ref{n2.P.NZ}) describes the dynamics accurately for all $T_\textrm{Z} \ll t$, which is shown in Fig. \ref{fig:n2.surv.prob}(b) (see also Ref. \cite{GPSS13} for discussion relevant to this point as well as the influence of a virtual bound state on the power law decay).
%This influence of the virtual bound states on the power law decay appears to be fairly general \cite{GPSS13}.
We can quantify the divergence of the timescale $T_\Delta$  in terms of the distance $\delta$ from the transition point $g=1$ after reparameterizing according to $g \equiv 1 - \delta$; then the timescale diverges like $T_\Delta \sim 1/\delta^2$ as $\delta \rightarrow 0$.

%%%
\begin{figure}
%\hspace*{0.01\textwidth}
 \includegraphics[width=0.45\textwidth]{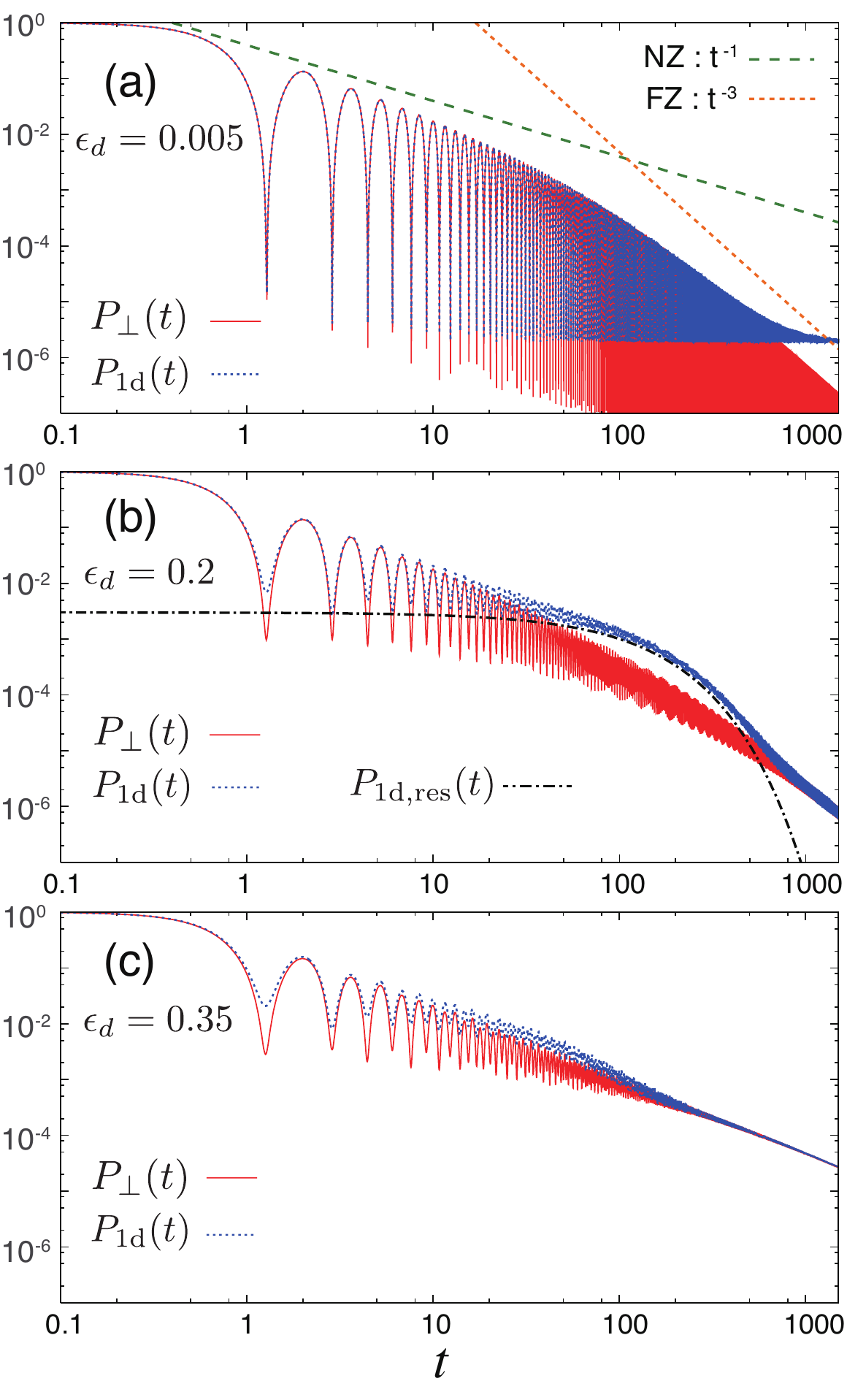}
\\
\vspace*{\baselineskip}
\caption{Numerical simulations for the survival probability and the non-escape probability for detuning from the BIC for $g=0.9$ and (a) $\epsd = 0.005$, (b) $\epsd = 0.2$, and (c) $\epsd =0.35$.
(Time $t$ is measured in units $1/J$ in which $J=1$.)}
\label{fig:4} 
\end{figure}
%%%

While the preceding analysis gives a clear picture of the types of evolution we can expect for the state 
$| \psi_\perp \ket$, it is still a bit idealized in comparison to experiment in two ways that we will account for below.  First, in a real experiment it would be difficult to tune exactly to the BIC at $\epsd = 0$; since the BIC is just the special case of a resonance with zero decay width, as we introduce detuning $\epsd \neq 0$ the resonance must reappear, which we could expect might perturb the non-exponential evolution of $P_\perp (t)$.  The complex eigenvalue of the resonance state can be expanded in the vicinity of the BIC up to second order in $\epsd$ as $z_\textrm{res} \approx \epsd / (1+g^2) - i \Gamma/2$ with $\Gamma = 2 g^2 \epsd^2 / (1+g^2)^3$, which of course reduces to $z_\textrm{BIC} = 0$ in the limit $\epsd = 0$.  However, when we examine $P_\perp (t)$ (red curve in Fig. \ref{fig:4} for $g=0.9$, as an example), we find that the resonance has virtually no influence on the survival probability, even for moderately large detuning values $\epsd \neq 0$.  We can obtain an understanding for this by calculating the resonance pole contribution to 
$P_\perp (t)$. Performing first a simple calculation for the pole contribution to the amplitude 
$\bra \psi_\perp | e^{-iHt} | \psi_\perp \ket$ reveals that, due to the geometric shape of the BIC-orthogonal state, both the lowest order and next-lowest order contributions in $\epsd$ cancel out, which yields
\beq
  P_{\perp,\textrm{res}} (t)
  	\approx \frac{g^4 \epsd^4}{\left(1+g^2\right)^8} e^{- \Gamma t}
%	\approx \frac{g^4 \epsd^4}{\left(1+g^2\right)^8} e^{- 2 \frac{g^2 \epsd^2}{(1+g^2)^3} t}
	.
\label{P.perp.res}
\eeq
The pre-factor in this expression, which is fourth order in $\epsd$, assures that the exponential effect will be quite small for almost any $\epsd \simeq 0$ regardless of the value of $g$.  For example, even for modest detuning $\epsd = 0.2$ and $g=0.9$ in Fig. \ref{fig:4}(b) [red curve], we have $g^4 \epsd^4 / \left(1+g^2\right)^8 \sim 10^{-5}$.

Second, while preparation of the initial state $| \psi_\perp \ket$ seems feasible, measuring the precise output state $\bra \psi_\perp |$ might prove more challenging.  
%We expect that either $|\bra \textrm{d} | e^{-iHt} | \psi_\perp \ket|^2$ or $|\bra 1 | e^{-iHt} | \psi_\perp \ket|^2$ should be more accessible quantities.  While it can be shown that either of these quantities individually captures the important features discussed below, we choose to focus on the quantity
Instead, it may be more realistic to consider the quantity
\beq
  P_{1\textrm{d}} (t)
  	\equiv |\bra 1 | e^{-iHt} | \psi_\perp \ket|^2 + |\bra \textrm{d} | e^{-iHt} | \psi_\perp \ket|^2
	,
\label{P.1d}
\eeq
which is equivalent to the {\it non-escape probability} that has appeared in the literature previously \cite{Muga_review2,Muga95,GCMM95,Cavalcanti98,GCMV07}.  
It can easily be shown that $P_{1\textrm{d}} (t) = P_\perp (t)$ for the case $\epsd = 0$, and hence all of our preceding detailed analytical results still apply directly at the BIC.  As shown in Fig. \ref{fig:4}, the difference between $P_{1\textrm{d}} (t)$ [blue curve] and  $P_\perp (t)$ [red curve] appears first well into the long time region for small $\epsd \neq 0$ and moves gradually to earlier times as we increase the detuning.  The origin of the difference between the two quantities is easy to understand 
as it seems to be entirely attributable to the fact that only the lowest-order contribution in $\epsd$ cancels out when we calculate the resonance pole contribution to the amplitude for the non-escape probability $P_{1\textrm{d}} (t)$.  In particular, we find
\beq
  P_{1\textrm{d},\textrm{res}} (t)
  	\approx \frac{g^2 \epsd^2}{\left(1+g^2\right)^4} e^{- \Gamma t}
	,
\label{P.1d.res}
\eeq
which is still small, but has some noticeable influence on the spectrum in some cases.  For example, in Fig. \ref{fig:4} (b) for $\epsd = 0.2$ we see the resonance pole with magnitude $g^2 \epsd^2 / \left(1+g^2\right)^4 \sim 0.003$ introduces exponential dynamics into $P_{1\textrm{d}} (t)$ around $t \gtrsim 10$, although this only lasts for a few lifetimes $\tau = 2 / \Gamma \sim 360$, which leaves the non-escape probability relatively intact when this quantity rejoins with $P_\perp (t)$ as the $1/t^3$ far zone dynamics kick in.  We note that $P_{1\textrm{d}} (t)$ also exhibits the interesting feature of {\it pre}-exponential decay that extends beyond the usual parabolic dynamics in the region $1 \lesssim t \lesssim 10$.  As we further increase $\epsd$ as in Fig. \ref{fig:4} (c), we find the exponential decay region lasts even fewer lifetimes as the difference between $P_{1\textrm{d}} (t)$ and $P_\perp (t)$ again becomes diminished. 

In this work we have shown that by populating a state that lies orthogonal to a bound state in continuum one can observe non-exponential dynamics that are usually overwhelmingly suppressed when the resonance condition is satisfied.  
%We further showed that this effect is robust against even significant parametric detuning from the BIC.
Note that for the present model we could consider the evolution of more general BIC orthogonal states such as 
$g | d \ket + | 1 \ket + \sum_{n=2}^{\infty} w_n | n \ket$ that include elements of the chain beyond the BIC sector.  We briefly comment on a representative example of this more general configuration in Ref. \cite{Supp}, where we show that including a single site from the chain can suppress oscillations in the survival probability.

%%
%When we introduced the output measurement quantity $P_{1\textrm{d}} (t)$ that we believe lies closer to real experiment, the usual exponential decay reappears for some values of the system parameters, but fails to overwhelm the non-exponential dynamics as is usually the case.  Note from a certain point of view, $P_{1\textrm{d}} (t)$ might actually be considered the more interesting quantity to measure, recalling that our original motivation was to more easily observe {\it post-exponential} decay.  We have demonstrated our idea using a model based on the optical waveguide array experiment in Ref. \cite{BIC_opt_expt2} with a semi-infinite geometry.  Note one could perform a similar analysis for the optical waveguide array with infinite geometry and two side-attached `impurity' elements in Ref. \cite{BIC_opt_expt1}; unfortunately, in this case one would have to deal with two ordinary bound states that would persist outside the continuum edges even for impurity energies embedded deeply within the continuum \cite{GNHP09,TGKP16}, which would tend to somewhat obscure the interesting non-exponential dynamics.  However, for sufficiently small coupling the influence of these bound states could perhaps be minimized.  While these experimental proposals reside in the realm of optics, an interesting extension of the present work could be to develop a true quantum experiment in which to observe the non-exponential dynamics.

We briefly note we have focused here on bound states in continuum that appear purely due to interference effects as originally proposed by von Neumann and Wigner in 1929 \cite{BIC_1929}.  We have not directly addressed ``accidental'' BICs \cite{Hsu13} that exhibit interesting topological properties \cite{BIC_review,BIC_lasing,BM17,ZHLSS14}, although the study of BIC-orthogonal states in this context might prove fruitful as well.

%We point out a different approach for studying non-Markovian dynamics involving a BIC appears in \cite{XCLYG14}, focusing instead on the short-time Zeno dynamics.

%%%
%\begin{figure}
%\hspace*{0.01\textwidth}
% \includegraphics[width=0.45\textwidth]{fig4}
%\\
%\vspace*{\baselineskip}
%\caption{Time evolution for detuning from the BIC with $g=1$ and (a) $\epsd = 0.05$ and (b) $\epsd = 0.2$.}
%\label{fig:3} 
%\end{figure}
%%%

%%%
%\begin{figure}
%\hspace*{0.05\textwidth}
% \includegraphics[width=0.4\textwidth]{fig1a}
%\hfill
% \includegraphics[width=0.4\textwidth]{fig1b}
% \hspace*{0.05\textwidth}
%\\
%\vspace*{\baselineskip}
%\hspace*{0.01\textwidth}(a)\hspace*{0.47\textwidth}(b)\hspace*{0.4\textwidth}
%\\
%\caption{(color online) blah blah}
%\label{fig:contC} 
%\end{figure}
%%%

\section*{Acknowledgements}

The authors wish to thank M. V. Berry, J. G. Muga and K. Nishidai for helpful discussions and encouragement related to this project.  This work was supported in part by Japan Society for the Promotion of Science KAKENHI Grants No. JP18K03466 and JP16K05481.  S. G. also acknowledges support from the Research Foundation for Opto-Science and Technology.  D. S. acknowledges support from the Canada Research Chair Program.

%%%%%%%%%%%%%
%%%%%%%%%%%%%

%%%%%%%%%%%%
%%%%%%%%%%%%

\pagebreak

\renewcommand{\thepage}{S\arabic{page}}
\renewcommand{\thesection}{S\arabic{section}}
\renewcommand{\thefigure}{S\arabic{figure}}
\renewcommand{\theequation}{S\arabic{equation}}

\setcounter{page}{1}
\setcounter{equation}{0}
\setcounter{figure}{0}

%\section{Supplementary Material}

%%%%%%%
\section{Supplementary Material: Derivation of resolvent operator and self-energy}
\label{sec:resolvent}

To obtain the explicit expression for the resolvent operator, we first rewrite the Hamiltonian from the main text as $H = H_0 + V$, in which
\beq
  H_0 = \epsd | d \ket \bra d | + \int_0^\pi dk \; E_k | k \ket \bra k | 
\eeq
and
\beq
  V = g \int_0^\pi dk \; V_k \left( | d \ket \bra k | + | k \ket \bra d | \right) 
  	.
\eeq
Then after applying a simple operator expansion
\beqa
  \bra d | \frac{1}{z - H} | d \ket
    &	= & \bra d | \left( \frac{1}{z - H_0} + \frac{1}{z - H_0}V \frac{1}{z - H_0} \right.
					\nonumber  \\
    &	& + \left.  \frac{1}{z - H_0}V \frac{1}{z - H_0}V \frac{1}{z - H}  \right) | d \ket
	,
\eeqa
we can easily solve for the explicit form of the resolvent operator 
$\bra d | (z - H)^{-1} | d \ket = (z - \epsd - \Sigma (z))^{-1}$ given in the main text (note the second term of the above expansion vanishes).

Our next task is to perform the necessary integration to obtain the explicit form of the self-energy function 
$\Sigma(z)$.  This can be achieved through a variety of methods; for example, by applying the integration transformation $w=e^{ik}$ we can write
\beqa
  \Sigma (z)
     &	= & g^2 \int_0^\pi \frac{| V_k |^2}{z - E_k} dk
	= \frac{g^2}{\pi} \int_{- \pi}^{\pi} dk \frac{\sin^2 2k}{z + 2 \cos k} 
				\nonumber  \\
	%%
%    &	= & - \frac{g^2}{4 \pi i} \oint_\Lambda dw \frac{w^4 -2 + w^{-4}}{w^2 + z w + 1}
%    				\nonumber  \\
	%%
    &	= & - \frac{g^2}{2 \pi i} \oint_\Lambda dw \frac{w^4 - 1}{(w - w_1)(w - w_2)}
\label{SM.sigma.z.int}
\eeqa
in which $\Lambda$ is the counter-clockwise contour just inside the unit circle in the complex $w$-plane and $w_{1,2} = (-z \pm \sqrt{z^2 - 4} )/2$.
%and we have transformed the integration variable according to $\bar{w} = w^{-1}$ in the third term of the second line in order to obtain the expression reported in the third line.  
%To make the final evaluation of the integral we write $w^2 + z w + 1 = (w - w_1)(w - w_2)$ in which $w_{1,2} = (-z \pm \sqrt{z^2 - 4} )/2$.  
Since $w_1 w_2 = 1$, if $w_1$ satisfies $|w_1| < 1$ then we must have $|w_2| > 1$ (or vice versa) and hence exactly one solution always falls inside the unit circle (we treat the situation $|w_1| = |w_2| = 1$ as a limit of the other two cases).  
Evaluating the integral as a residue then yields the expression for the self-energy reported in the main text, where the sign in front of the square root is a $-$ ($+$) whenever the solution $w_1$ ($w_2$) appears inside the unit circle.  %\textcolor{red}{Note this expression is consistent with the literature for the present model (although the notation for dealing with the sign change and the analytic continuation for complex $z$ vary somewhat) \cite{LonghiEPJ07,EG10,Fukuta}.}

For the case of a bound state satisfying $z>2$, one can show the $|w_1| < 1$ case holds.
Taking the residue in Eq. (\ref{SM.sigma.z.int}) then yields the expression for the self-energy reported in %Eq. (4) of 
the main text.  
As an explicit example, consider the upper bound state $z_+$ appearing in the case $\epsd = 0$ and $g > 1$ from the main text.  With $z_+ = g + 1/g$ we find $w_1 (z_+) = - 1/g$, so that $-1 < w_1 < 0$ inside the unit circle in the complex $w$-plane, as expected.
Note that we can also determine the wave vector $k_+$ that appears in the associated wave function $\bra x | \psi_+ \ket \sim e^{i k_+ x}$ from the dispersion relation $z_+ = - 2 \cos k_+$. %or %$g + 1/g = -e^{i k_+} - e^{-i k_+}$, 
%%
%\beq
%  g + \frac{1}{g} = -e^{i k_+} - e^{-i k_+}
%  	.
%\eeq
%%
Taking $e^{i k_+} = - 1/g$ gives $k_+ = \pi + i \log g$ with $\im k_+ > 0$ for $g>1$, which indeed yields a localized wave function.  This verifies that $z_+$ is indeed a bound state eigenvalue in this case, residing in the first Riemann sheet of the complex $z$ plane.
%(and hence resides on the first Riemann sheet in the complex $z$ plane, consistent with the literature).
%%
For the case $g < 1$, we are forced to analytically continue this solution into the second Riemann sheet as $w_1 < -1$ passes outside the unit circle.
Instead, we now have $-1 < w_2 < 0$ so that $w_2 = -g$ is the pole appearing inside the contour integration of Eq. (\ref{SM.sigma.z.int}), resulting in a sign change for the non-analytic part of the self-energy for the solution $z = z_+$.
The wave function associated with this eigenvalue now becomes divergent as the wave vector is most naturally written in this case as $k_+ = \pi - i \log \bar{g}$ with $\bar{g} = g^{-1} > 1$ and $\im k_+ < 0$.

One can perform a similar analysis for the lower (virtual) bound state satisfying $z_- < -2$, except that $w_1$ and $w_2$ switch roles compared to the above explanation; this implies the sign in front of the non-analytic part of the self-energy is reversed compared to scenario for the upper bound state $z_+$.  Note this requires that the sign designation for the lower bound state is opposite that of the upper bound state, without performing analytic continuation into the second Riemann sheet.  (This point can also be shown independently from the present analysis by working directly with the expression reported for the self-energy in the main text.)  Finally, the wave vector in this case is given by $k_- = i \log g$.

We note this expression for the self-energy from the main text has appeared in the literature previously \cite{LonghiEPJ07,Fukuta}.

%%%%%%%
\section{Details of approximate dynamics}
\label{sec:derivation}

%\begin{widetext}

%%%
\begin{figure*}
%\hspace*{0.01\textwidth}
 \includegraphics[width=0.85\textwidth]{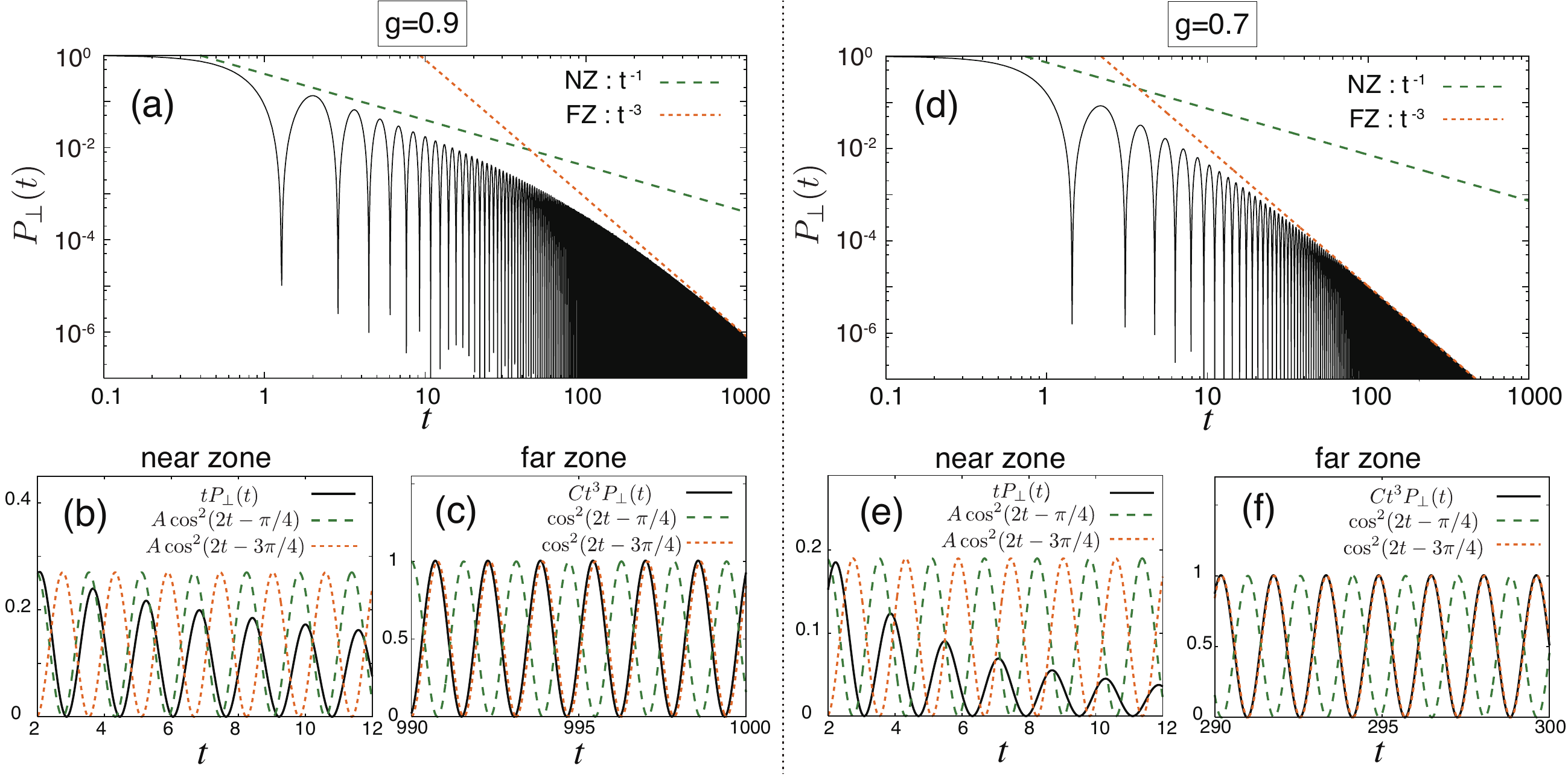}
\\
%\vspace*{\baselineskip}
\caption{Survival probability of $|\psi_\perp \ket$ at time $t$ for $\epsd = 0$ and
(a-c) $g=0.9$ and (d-f) $g=0.7$. (Compare with Fig. 2 from the main text.
Note $t$ is measured in units $1/J$ in which $J=1$.)
}
\label{fig:NZ.FZ} 
\end{figure*}
%%%

%\end{widetext}

We start with the exact expression for the dynamics associated with the branch cut taken from the main text
\beq
  A_\textrm{br} (t)
  	= \frac{1+g^2}{4 \pi i g^2} \int_{\mathcal{C}_\textrm{br}} dz e^{-izt} \frac{\sqrt{z^2 - 4}}{z^2 - z_g^2}
	.
\label{SM.A.br.defn}
\eeq
In Sec. \ref{sec:derivation.NZ} we outline the approximations for the short/intermediate time region and make a brief comment on the influence of the virtual Rabi oscillation on the phase in the near zone, while in Sec. \ref{sec:derivation.FZ} we describe corresponding approximations for the far zone.  In Sec. \ref{sec:derivation.plots} we present some plots for additional $g$ values compared to the main text and briefly discuss these.  Note that $\epsd = 0$ for all simulations in this document.

%%%%%%%
\subsection{Short/intermediate time region}
\label{sec:derivation.NZ}

We begin the derivation by performing a fraction decomposition on the integrand of Eq. (\ref{SM.A.br.defn}) in order to rewrite this as
\beq
  A_\textrm{br} (t)
  	= - \frac{1}{2g} \left( I(z_+) - I(z_-) \right)
\label{SM.A.NZ.1}
\eeq
in which 
\beq
  I (z_n) 
  	\equiv - \frac{1}{2 \pi} \int_{\mathcal{C}_\textrm{br}} dz e^{-izt} \frac{\sqrt{1 - z^2/4}}{z - z_n}
	.
\label{SM.A.NZ.I}
\eeq
From this point, we can apply methods similar to those appearing in Apps. C and D of Ref. \cite{GO17} to evaluate this integral; we eventually obtain
\beq
  I (\pm z_g) 
  	= e^{\mp i z_g t} \left[ \mp g - i \int_0^t d\tau e^{i z_g \tau} \frac{J_1 (2 \tau)}{\tau} \right]
	,
\label{SM.A.NZ.2}
\eeq
in which the first term is a pole contribution associated with the virtual bound states.  %Since we are primarily interested in the  vicinity of $g = 1$, %
In the short/intermediate time region delimited by $\Delta_g t \ll 1$ ($t \ll T_\Delta$), we can approximate the integral as
\beq
  I (\pm z_g) 
  	\approx \pm e^{\mp i z_g t} \left[ 1 - g - e^{\pm 2it} \left( J_0 (2t) \mp i J_1 (2t) \right) \right]
	,
\label{SM.A.NZ.3}
\eeq
Plugging this result into Eq. (\ref{SM.A.NZ.1}) gives
\beqa
  A_\textrm{br} (t)
  	& \approx & \frac{1}{g} 
		\left[ \left( g-1 \right) \cos z_g t + \cos (\Delta_g t) J_0 (2t) \right.
					\nonumber		\\
	& &	\left. - \sin (\Delta_g t) J_1 (2t) \right]
	.
\label{SM.A.NZ.f}
\eeqa
After applying the approximation $\Delta_g t \ll 1$ again we obtain the result in the main text Eq. (12).

As mentioned in the main text, the virtual Rabi oscillation plays a role in facilitating the phase shift from the early near zone (with phase $\pi/4$)
to the far zone (with phase $3 \pi/4$).
For example, at time $t = g/4\pi * T_\Delta \sim T_\Delta/10$ in the near zone evolution the coefficient of the second term in Eq. (13) from the main text has exactly half the magnitude of that of the first term; in this moment, the effective phase of the two combined terms can be shown to be 
$\phi_{1/2} = \arctan (\sqrt{2}/(\sqrt{2} -1)) \approx 0.4093 \pi$, which indeed satisfies $\pi/4 < \phi_{1/2} < 3\pi/4$.

%%%%%%%
\subsection{Asymptotic time region (far zone)}
\label{sec:derivation.FZ}

In the case of the asymptotic time zone $t \gg T_\Delta$, we find it most convenient to evaluate the integral Eq. (\ref{SM.A.br.defn}) using methods similar to those used in Ref. \cite{GPSS13}.
We begin by dragging the contour $\mathcal{C}_\textrm{br}$ surrounding the branch cut out to infinity in the lower half of the complex energy plane.  After this, the only non-vanishing portions of the integration are the contours $A_\mp (t)$ running from the two branch points out to infinity in the lower half plane.  These portions are written as
\beq
  A_\mp (t)
  	  = \frac{1+g^2}{2 \pi i g^2} \int_{\mp 2}^{\mp 2 - i \infty} dz e^{-izt} 
	  				\frac{\sqrt{z^2 - 4}}{\left(z - z_- \right) \left(z - z_+ \right)}
	.
\label{A.br.mp}
\eeq
Applying an integration variable transform $s \equiv it (z \pm 2)$ yields
\beq
  A_\mp (t)
  	  = \frac{i \left(1+g^2\right) e^{\pm 2it }}{2 \pi g^2 t^2} 
	  		\int_{0}^{\infty} ds e^{-s} \frac{\sqrt{s^2 \mp 4ist}}
					{\mp \Delta_g \left( 2 + z_g \right) + 4i\frac{s}{t} \mp \frac{s^2}{t^2}}
	.
\label{A.br.mp.2}
\eeq
For very large $t$ the first term in the denominator is much larger than the other two terms, which can be safely neglected.  Performing the remaining simplified integration and combining $A_\pm$ we obtain the result reported for the far zone in Eq. (15) of the main text.

%\textcolor{red}{Note that we could also write an approximation for the near zone evolution ($T_Z \ll t \ll T_\Delta$) by assuming the middle term of the denominator of Eq. (\ref{A.br.mp.2}) is instead the dominant term.  Doing so yields an accurate description of the $1/t$ dynamics in the early near zone, but however misses the virtual Rabi oscillation entirely.  Hence the approach in Sec. \ref{sec:derivation.NZ} is superior for the near zone calculation in the present situation.}

%%%%%%%
\subsection{Near zone/far zone transition: plots for additional cases}
\label{sec:derivation.plots}

\begin{figure}
\hspace*{0.01\textwidth}
 \includegraphics[width=0.45\textwidth]{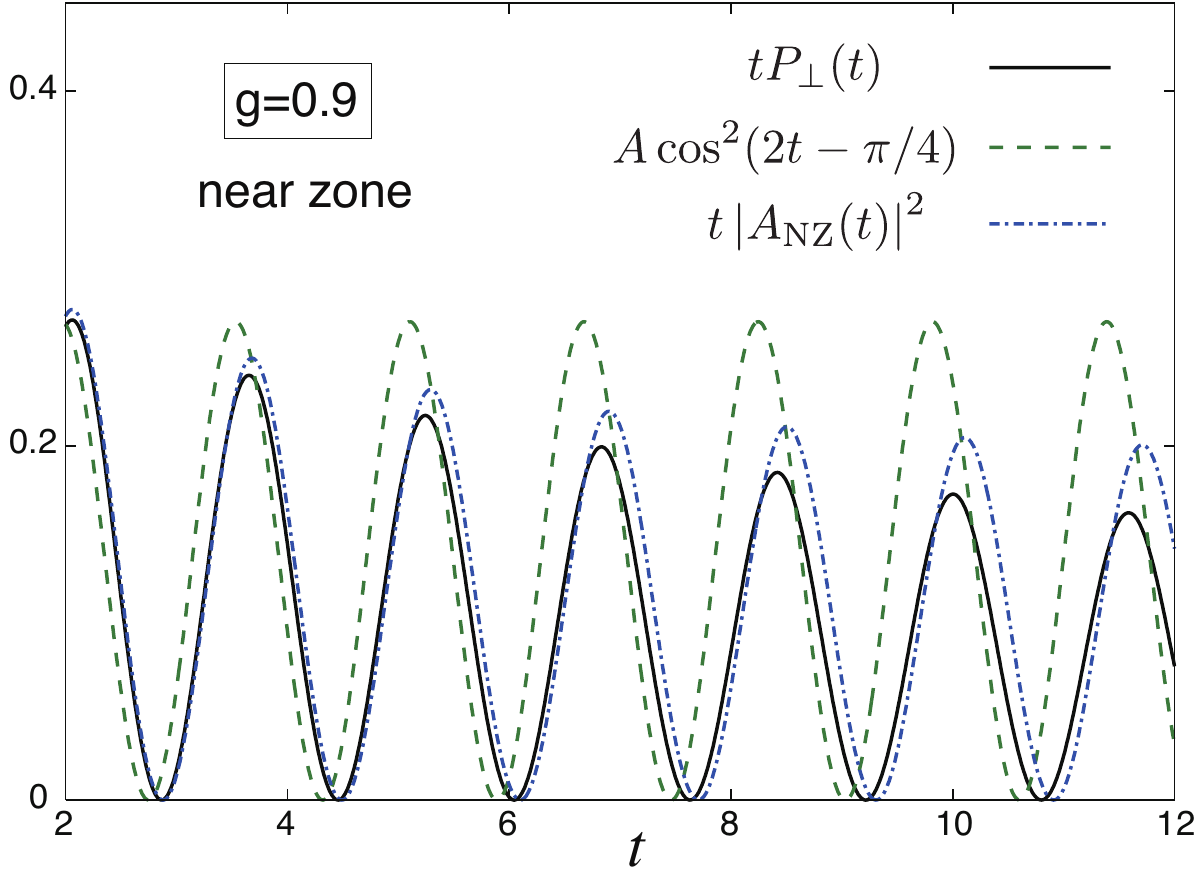}
\\
%\vspace*{\baselineskip}
\caption{Closer view of near zone plot from Fig. \ref{fig:NZ.FZ}(b), including comparison with analytic approximation resulting from including both terms of Eq. (13) [dashed-dotted curve] in the main text.
(Note time $t$ is measured in units $1/J$ in which $J=1$.)
}
\label{fig:NZ.close} 
\end{figure}

In Fig. \ref{fig:NZ.FZ}(a-c) we plot the survival probability $P_\perp (t)$ for $g=0.9$, similar to the case $g=0.98$ that was presented in Fig. 2(c-e) of the main text; only here we are a bit further away from the localization transition at $g=1$.
We see in Fig. \ref{fig:NZ.FZ}(b) for this case that the early near zone $1/t$ prediction gives only a rough description in terms of the amplitude of $P_\perp (t)$; however, the phase prediction $\cos^2 (2t - \pi/4)$ is still accurate.  We can improve our approximation for the amplitude by including the second term from Eq. (13) in the main text, which is shown explicitly as the dashed-dotted curve in Fig. \ref{fig:NZ.close}. 
We can estimate the point $T_\textrm{br}$ at which this approximation, too, begins to breakdown as about 10\% of $T_\Delta$.  For $g=0.9$ we find this occurs around $t \approx 9$, in rough agreement with Fig. \ref{fig:NZ.close}.

We plot the same in Fig. \ref{fig:NZ.FZ}(d-f) for $g=0.7$, significantly further from the $g=1$ localization transition.  In this case, our analytic near zone approximation breaks down, as 10\% of $T_\Delta$ occurs at about $t \approx 0.7$, before the near zone dynamics even emerge.  However, we still achieve our primary objective of complete non-exponential decay.

If we keep decreasing the value of $g$, eventually around $g \approx 0.38$ we obtain $T_\Delta \sim 1$.  For this and any smaller values of $g$ the near zone is entirely squeezed out and the system will instead transition from the early time parabolic (Zeno) dynamics directly into the $1/t^3$ far zone decay.  But again, the evolution is still entirely non-exponential.

We comment that all numerical results in this work were obtained by evolving a chain in the site representation (with up to 16000 elements) according to the Schr\"odinger equation using a variable-order variable-step Adams method.

%solving a coupled differential equations for expansion coefficients in the site representation, which is directly derived from the time-dependent Schrodinger equation, using a variable-order variable-step Adams method. The number of sites was taken as a large number up to 16000 to avoid artificial reflections due to truncation of the equations.

%%%FIG
%%%
%%%

%%%%%%%
\section{Chain-induced effective decoherence}
\label{sec:gen}

\begin{figure}
\hspace*{0.01\textwidth}
 \includegraphics[width=0.45\textwidth]{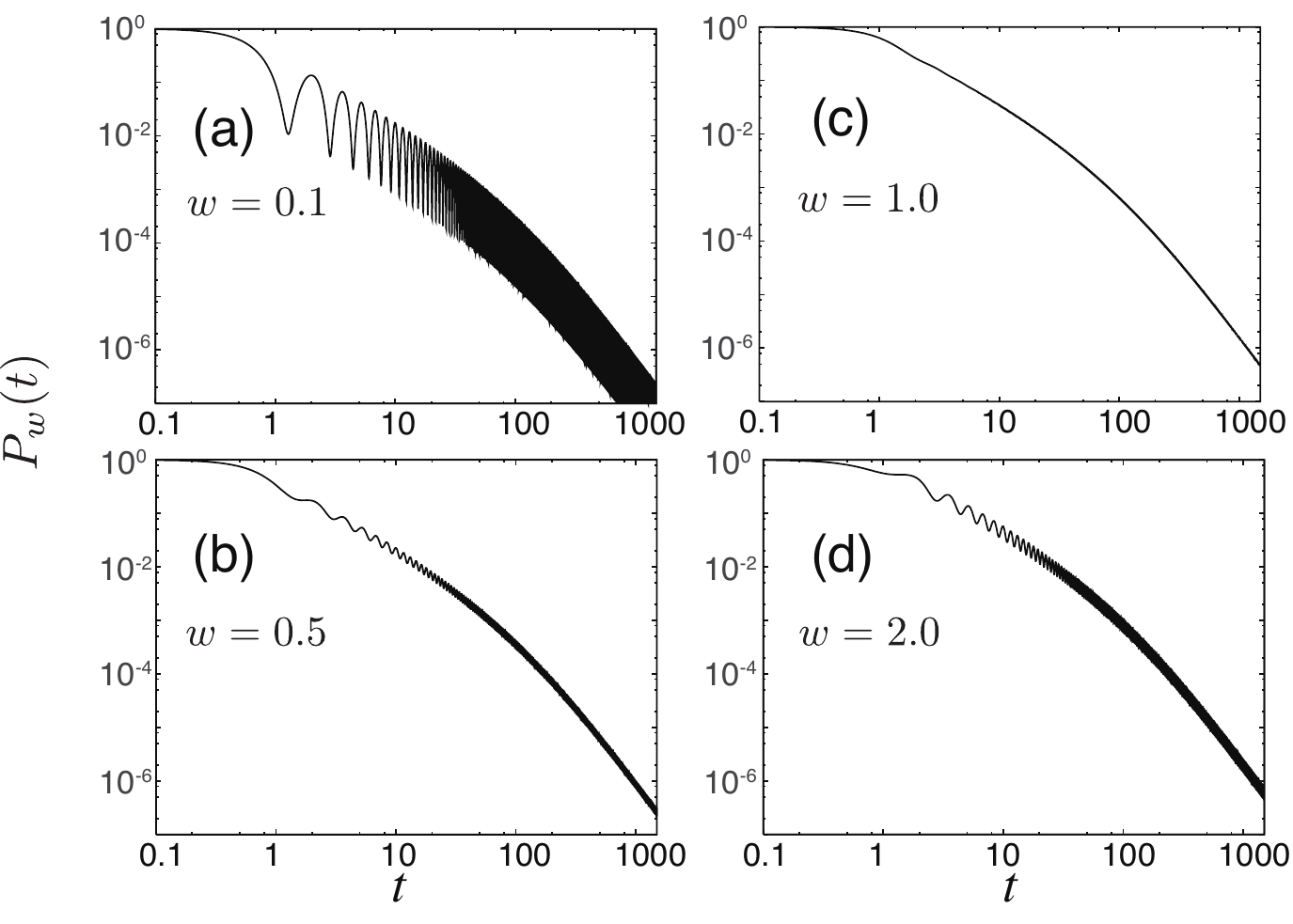}
\\
%\vspace*{\baselineskip}
\caption{Numerical simulations for the survival probability for the $| \psi_w \ket$ state with $g = 0.9$, $\epsd=0$ and (a) $w=0.1$, (b) $w=0.5$, (c) $w=1.0$, and (d) $w=2.0$.
(Note $t$ is measured in units $1/J$ in which $J=1$.)
}
\label{fig:w} 
\end{figure}

Here we briefly consider the evolution of a slightly more general BIC orthogonal state, written as
\beq
  | \psi_w \ket
  	= N_w \left( g | d \ket + | 1 \ket + w | 2 \ket \right)
\label{psi.w}
\eeq
in which $N_w^2 = (1 + g^2 + w^2)^{-1}$.  Here we have included a single site $| 2 \ket$ from the chain outside of the subspace spanned by the BIC itself, with amplitude $w$.
In Fig. \ref{fig:w} we show how the inclusion of this site modifies the evolution for $g=0.9$.  In Fig. \ref{fig:w} (a-c), we see that increasing the value of $w$ in the range $w \le 1$ results in the oscillations we observed in the main text becoming gradually damped out, with near total damping occurring for $w=1$. By contrast, the oscillations return for $w$ values much larger than 1 as shown in Fig. \ref{fig:w} (d).

%%%FIG
%%%
%%%

\begin{figure}
\hspace*{0.01\textwidth}
 \includegraphics[width=0.45\textwidth]{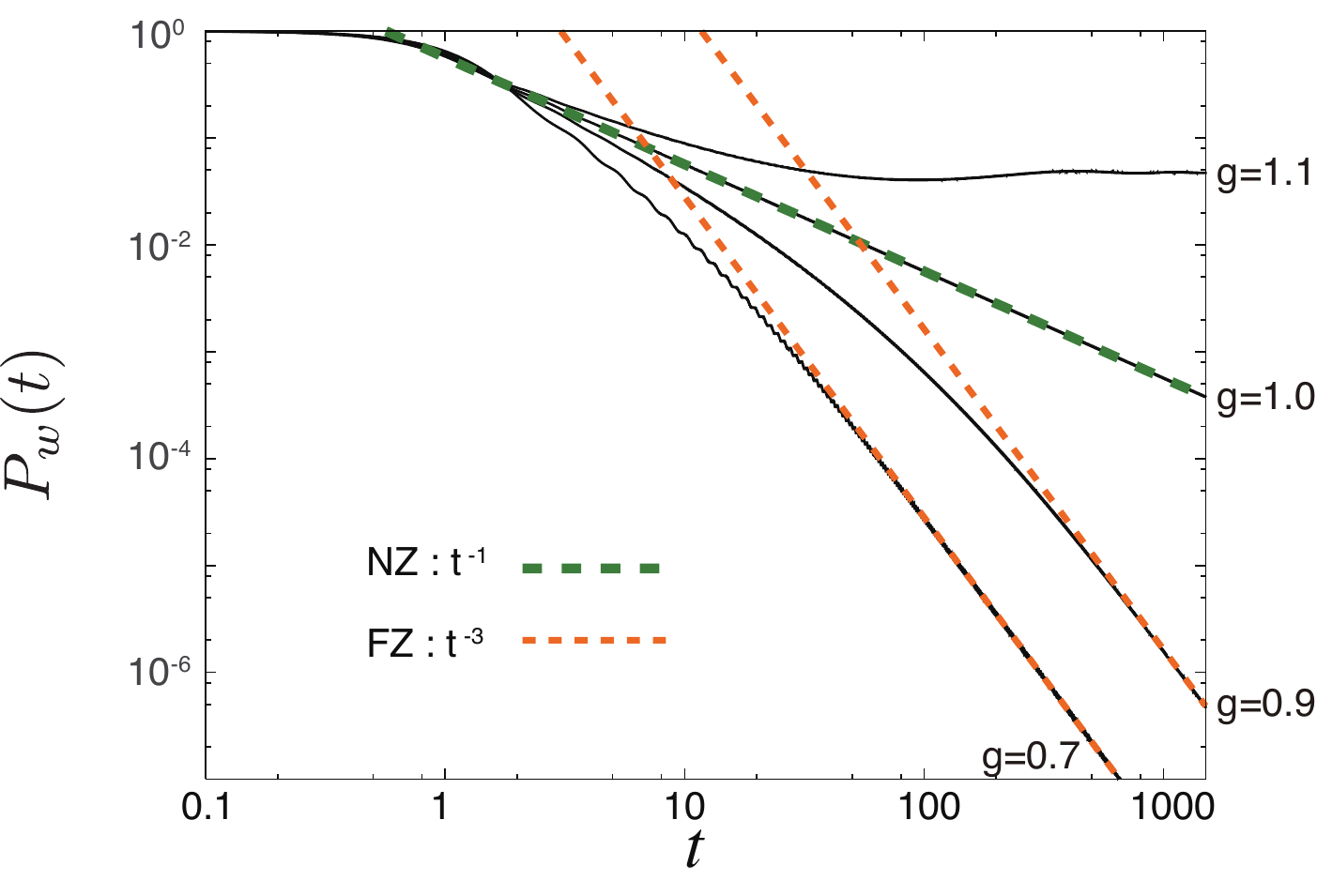}
\\
%\vspace*{\baselineskip}
\caption{Numerical simulations for the survival probability for the $| \psi_w \ket$ state for $w=1$ and $\epsd=0$ with various values of $g$.
Predictions for the near zone at $g=1$ [Eq. (\ref{P.w.NZ.1}), dashed line] and the far zone [Eq. (\ref{P.w.FZ}), dotted line] are indicated.  (Note values $g=1.1$ and $g=1$ compare with Fig. 2 (a) and (b), respectively, in the main text.  Also note time $t$ is measured in units $1/J$ in which $J=1$.)
}
\label{fig:w_1} 
\end{figure}

Extending from this observation, in Fig. \ref{fig:w_1} we plot numerical simulations for the survival probability of $| \psi_w \ket$ for $w=1$ over a variety of $g$ values.  We observe that the near-total suppression of the oscillations occurs for a wide range of $g$ values in the vicinity of $g=1$.

%%%
%%%

We can gain insight into the mechanism of this suppression through the following analytic approximations.
We begin by writing the survival probability for this state
$P_w (t) = |A_w (t)|^2$, in which
\beqa
  A_w (t)
     &	= & \bra \psi_w | e^{-iHt} | \psi_w \ket
	= \frac{1}{2 \pi i} \int_{\mathcal{C}_E} dz \; e^{-izt} \bra \psi_w | \frac{1}{z - H} | \psi_w \ket
						\nonumber	\\
     &	= & \frac{N_w^2}{2 \pi i} 
		\int_{\mathcal{C}_E} dz \; e^{-izt} \left( \sigma_1 (z) + Q(z) G_\textrm{dd}(z) \right)
	.
\label{A.w.defn}
\eeqa
Here we have
\beq
  \sigma_1 (z)
  	= \frac{z - \sqrt{z^2 - 4}}{2}
	,
\label{sigma.1}
\eeq
\beqa
  Q (z)
    & = & g^2 + \left[ \sigma_1 (z) \right]^2 \left( 2g^2 - 2g^2 wz \right.
				\nonumber 	\\
    & &	\left. + g^2 \left[ \sigma_1 (z) \right]^2  -2wz +w^2 z^2 \right)
	,
\label{Q.z}
\eeqa
and $G_\textrm{dd}(z) = \left( z - \epsd - \Sigma (z) \right)^{-1}$ is the resolvent operator at the impurity site, Eq. (3) from the main text.

Focusing on the case $w = 1$ as a quick example, it can be shown that $Q(z)$ simplifies a bit as
\beq
  Q(z)
  	= \left( 1 + g^2 \right) z \left( z - 2 \right) \left[ \sigma_1 (z) \right]^2
	.
\label{Q.z.1}
\eeq
Note the useful relation $\sigma_1 (z) + 1/\sigma_1 (z) = z$ has been applied here.  Eq. (\ref{Q.z.1}) can in turn be used to simplify the integrand of Eq. (\ref{A.w.defn}) such that we obtain
\beq
  A_w (t) 
  	= \frac{N_w^2}{4 \pi i} 
		\int_{\mathcal{C}_E} dz \; e^{-izt} \frac{\left( z - g z_g \right)^2 \sqrt{z^2 - 4} }
					{\left( z - z_g \right) \left( z + z_g \right)}
	.
\label{A.w.1}
\eeq
Note the presence of the $(z - g z_g)^2$ factor in the numerator of the integrand.  Since the dominant contribution to the integration comes from around the branch points $z= \pm 2$ and since $g z_g = 1 + g^2 \approx 2$ in the vicinity of $g \sim 1$, this factor is very small for the contribution coming from the upper branch cut (if we had chosen $w = -1$, it would instead be the lower branch cut contribution that would be very small).  Since there is one overwhelmingly dominant contribution, the oscillations between the two band edges are greatly diminished, which explains the effective decoherence observed in Figs. \ref{fig:w} and \ref{fig:w_1}.

To see this more explicitly, we carry out the integration for the far zone in the $g \neq 1$ case by dragging the integration contour out to infinity in the lower half plane similar to Sec. \ref{sec:derivation.FZ}.  Doing so we find there are again the two contributions $A_w (t) = A_{+,w} (t) + A_{-,w} (t)$ in which
\beq
  A_{\mp,w} (t)
	= \mp \frac{N_w^2 \left( \mp 2 - g z_g \right)^2 e^{\mp i (2t - \pi/4)} }
			{2 \sqrt{\pi} g^2 \left( 2 + z_g \right) \Delta_g t^{3/2} }
	.
\label{A.w.2}
\eeq
Hence we immediately see the $A_+ (t)$ contribution will indeed be quite small for $g \sim 1$.
The far zone evolution is then approximately described by
\beq
  P_{\textrm{FZ},w} (t)
	\approx \frac{\left( 2 + g z_g \right)^4 }
			{4 \pi g^4 \left( 2 + g^2 \right)^2 \left( 2 + z_g \right)^2 \Delta_g^2 t^{3} }
	,
\label{P.w.FZ}
\eeq
which predicts the slope in Fig. \ref{fig:w_1} quite well (red dotted lines), even in the case $g=0.7$ not so near the localization transition at $g=1$.  [Compare this result with Eq. (15) in the main text.]

In the special case $g=1$ notice that $ z - gz_g = z - 2$ exactly, which results in the upper band edge contribution vanishing entirely.  The timescale 
$T_\Delta$ also diverges as the gap $\Delta_g$ closes (just as in the main text), which leads to an asymptotic near zone ($1/t$) description 
\beq
  P_w (t) \approx 16/ 9 \pi t
	.
\label{P.w.NZ.1}
\eeq
This again agrees very well (green dashed line) with the numerical simulation in Fig. \ref{fig:w_1}.   [Compare this result with Eq. (14) in the main text.]

%%%%%%%
\section{Brief comment on model geometry}
\label{sec:geo}

%%%
\begin{figure}
\hspace*{0.01\textwidth}
 \includegraphics[width=0.35\textwidth]{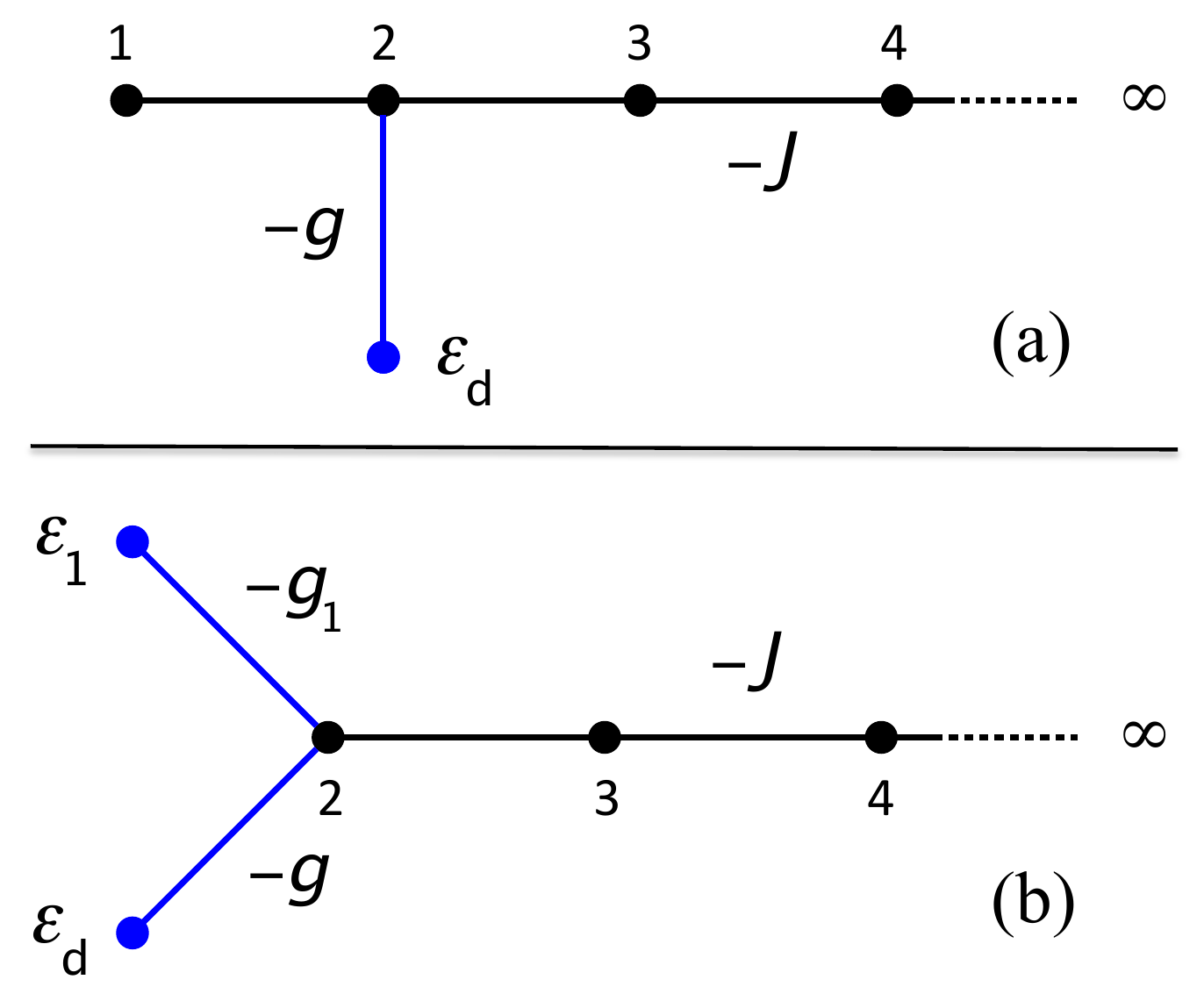}
\\
%\vspace*{\baselineskip}
\caption{
(a) Original geometry from the main text with a single side-coupled impurity $| d \ket$; (b) alternative geometry with two impurities, $| d \ket$ and $| 1 \ket$.  The second model reduces to the first for $g_1 = J$ and $\epsilon_1 = 0$. %$g_2 = g$ and $\epsilon_2 = \epsilon_\textrm{d}$.
}
\label{fig:geo} 
\end{figure}
%%%

In this work, we have relied on the model depicted in Fig. \ref{fig:geo}(a) to illustrate our idea about populating a BIC-orthogonal state to suppress the exponential decay process.  This model consists of a semi-infinite tight-binding chain (with sites $n = 1, 2, ... \infty$) coupled to an `impurity' (discrete) state $|d\ket$.  One might pause when considering that the initial state studied in this work (the BIC-orthogonal state) is a combination of the impurity $|d \ket$ and an element taken from the chain $| 1 \ket$, the latter of which is included in the environment portion of the Hamiltonian as it is written in Eq. (1) in the main text.

However, we emphasize here that viewing the $|1\ket$ state as part of the environment is rather arbitrary, as one could easily imagine a slightly more general model, consisting of a double impurity sector as illustrated in Fig. \ref{fig:geo}(b).  Here we have two impurity sites $|1\ket$ and $|d\ket$ where $|1\ket$ now has the generalized energy $\epsilon_1$ and is coupled to the chain with strength $-g_1$.  The semi-infinite chain now consists of sites $n = 2, 3 ... \infty$.  Clearly the double-impurity model (b) reduces to the original model (a) when $g_1 = J$ and $\epsilon_1 = 0$.  This illustrates that, at least from a theoretical perspective, whether site $|1\ket$ is viewed as part of the impurity sector or as part of the reservoir is arbitrary.

Throughout this work we chose to view the model as depicted in Fig. \ref{fig:geo}(a), in part for convenience and in part because that is how this model has previously been presented in the literature, as a specific case of the models in Refs. \cite{LonghiEPJ07,Fukuta}.  However, when performing the experiment proposed in the main text, it might be more natural to adopt the perspective from Fig. \ref{fig:geo}(b).  For example, we might view the states $\{ | d \ket, | 1 \ket \}$ as two modes of a single waveguide in a potential waveguide array experiment; the experimentalist then achieves the initial state $| \psi_\perp \ket$ by preparing a coherent superposition of the two modes.

\end{document}